\documentclass{article}

\pdfoutput=1

\usepackage[utf8]{inputenc}
\usepackage{xcolor}
\usepackage{graphicx}
\usepackage{chemformula}
\usepackage{xspace}
\usepackage{ulem}
\usepackage{amssymb}
\usepackage{amsmath}
\usepackage{hyperref}
\usepackage{mathtools}
\usepackage{authblk}

\definecolor{dgreen}{HTML}{008000}
\newcommand{\fig}[1]{Fig.~\ref{fig:#1}}
\newcommand{\rw}{$R_w$\xspace}
\renewcommand{\vec}[1]{\mathbf{#1}}

\begin{document}

\title{Refining perovskite structures to pair distribution function data using collective Glazer modes as a basis}

\author[a]{Sandra Helen Skj{\ae}rv{\o}}
\author[b]{Martin~A. Karlsen}
\author[c]{Riccardo Comin}
\author[a,d]{Simon~J.~L. Billinge}

\affil[a]{Department of Applied Physics and Applied Mathematics, Columbia University, New York, NY, 10027, USA}
\affil[b]{Department of Physics, Chemistry and Pharmacy, University of Southern Denmark, 5230, Odense M, Denmark}
\affil[c]{Physics Department, Massachusetts Institute of Technology, Cambridge, MA, 02139, USA}
\affil[d]{Condensed Matter Physics and Materials Science Department, Brookhaven National Laboratory, Upton, NY, 11973, USA}

\maketitle

\begin{abstract}
Structural modelling of octahedral tilts in perovskites is typically done using the symmetry constraints of the resulting space group. In most cases, this introduces more degrees of freedom than those strictly necessary to describe only the octahedral tilts. It can therefore be a challenge to disentangle the octahedral tilts from other structural distortions such as cation displacements and octahedral distortions. This paper reports on the development of constraints for modelling pure octahedral tilts and implemented the constraints in diffpy-CMI, a powerful package to analyse pair distribution function (PDF) data. The program allows features in the PDF that come from rigid tilts to be separated from non-rigid relaxations, provides an intuitive picture of the tilting, and as it has many fewer refinable variables than the unconstrained space-group fits, provides robust and stable refinements of the tilt components. It further demonstrates the use of the model on the canonical tilted perovskite \ch{CaTiO3} which has a known Glazer tilt system $\alpha^+ \beta^- \beta^-$. The Glazer model fits comparably to the corresponding space group model $Pnma$ below $r = 14$~\AA\ and becomes progressively worse than the space group model at higher $r$ due to non-rigid distortions in the real material. 
\end{abstract}

\section{Introduction}

\begin{figure*}[bh]
    \centering
    \includegraphics[width=0.4\textwidth]{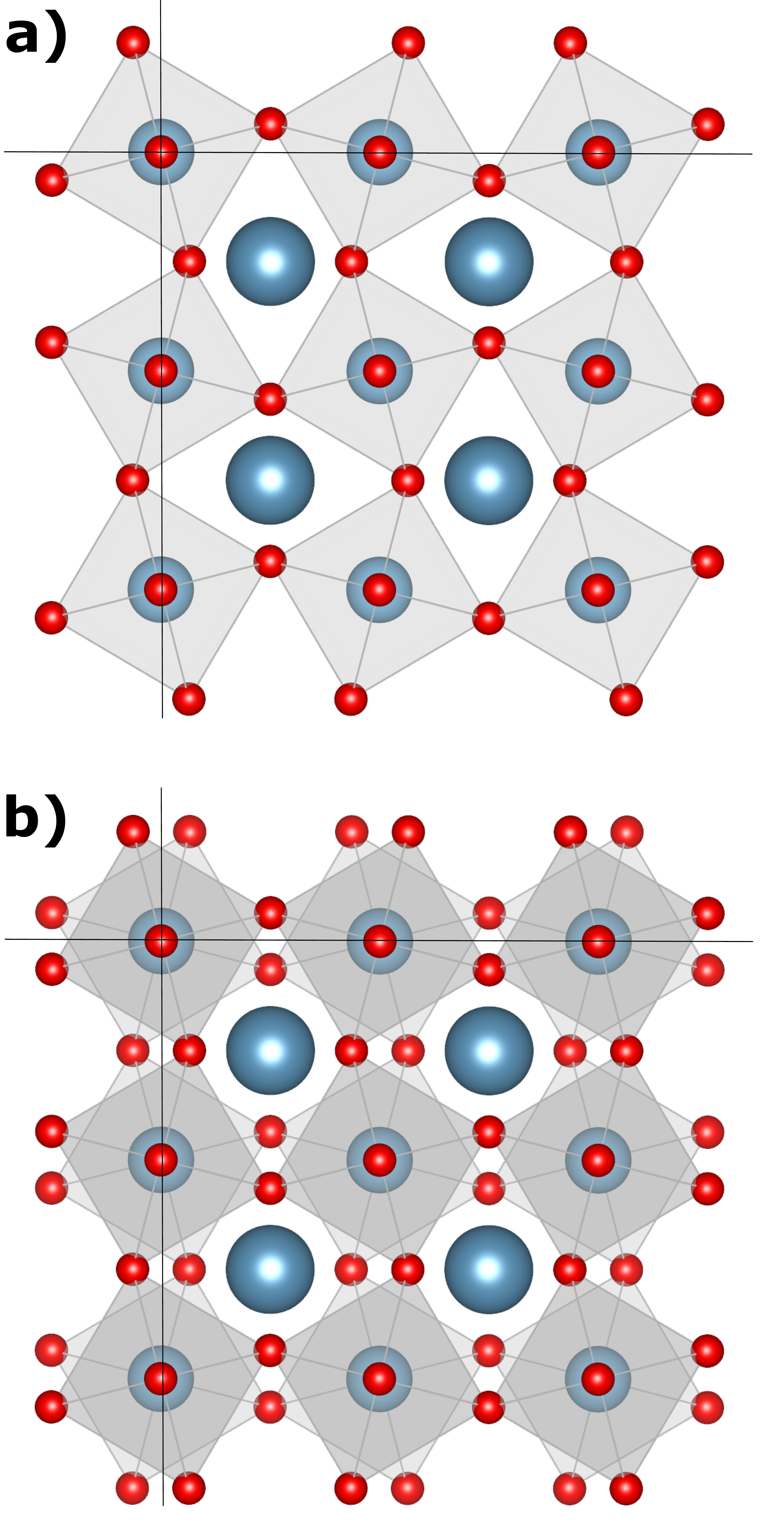}
    \caption{Illustration of in-phase and out-of-phase tilt systems as viewed down the tilt axis.  The tilt systems shown here are $\alpha^0 \alpha^0 \gamma^+$ (top) and $\alpha^0 \alpha^0 \gamma^-$ (bottom).}
    \label{fig:illustration-of-IP-OOP-tilts}
\end{figure*}

Structural distortions in materials, such as those occurring during displacive structural phase transitions, often involve collective displacements of groups of atoms \cite{doveTheoryDisplacivePhase1997}. 
For example, in the perovskites, a material class with nominal stoichiometry \ch{ABX3} (\fig{illustration-of-IP-OOP-tilts}), collective distortions are known to cause a host of structural phase transitions that lower the symmetry of the cubic parent structure \cite{mullerCharacteristicStructuralPhase1968, saljePhaseTransitionsFerroelastic1990, goodenoughTheoryRoleCovalence1955a, kweiStructuresFerroelectricPhases1993a, kweiPairdistributionFunctionsFerroelectric1995d}.

Distortions away from from the cubic archetype can involve deformations of the octahedra, displacements of the B cations inside the octahedra and tilting of the octahedra. The first two are typically caused by electronic instabilities, while the latter is due to the relative sizes of the cations. For perovskites with smaller A cations, the octahedra tilt to compress the structure around them, essentially improving the bonding for the A cation. This geometric effect is conveniently captured by the Goldschmidt tolerance factor \cite{goldschmidtGesetzeKrystallochemie1926}, 
\begin{equation}
    t=\frac{r_\mathrm{A} + r_\mathrm{X}}{\sqrt{2}(r_\mathrm{B} + r_\mathrm{X})}
\end{equation}
where $r$ is an ionic radius and subscripts A, B and X denote the ion type. For $t = 1$, the perovskite crystallizes in the high-symmetry cubic structure, while octahedral tilting is expected for a $t < 1$ as it signifies that the A site cation is too small to fill the void between the octahedra. In this paper we will concentrate on the latter type of distortion.

Due to their corner-sharing geometry, the octahedra can tilt collectively in several different patterns.  By building macroscopic models of corner-shared rigid octahedra, Glazer was able to describe all the 22 different patterns in which the rigid octahedra could collectively tilt, and the resulting symmetry space groups \cite{glazerClassificationTiltedOctahedra1972a, glazerSimpleWaysDetermining1975a}. Later studies uncovered details about these Glazer systems through group theory and geometric considerations \cite{aleksandrovSequencesStructuralPhase1976, okeeffeStructuresTopologicallyRelated1977, woodwardOctahedralTiltingPerovskites1997a, woodwardOctahedralTiltingPerovskites1997b, howardGroupTheoreticalAnalysisOctahedral1998}.

Depending on the Glazer tilt pattern of a perovskite, the structure will have different symmetry space group \cite{aleksandrovSequencesStructuralPhase1976, okeeffeStructuresTopologicallyRelated1977, woodwardOctahedralTiltingPerovskites1997a, woodwardOctahedralTiltingPerovskites1997b, howardGroupTheoreticalAnalysisOctahedral1998}. Modelling the structures of the these low-symmetry phases is therefore often done using the symmetry-broken crystallographic models and constraining the allowed atomic displacements to those imposed by the space-group symmetries.
However, in general, the symmetry space group allows for more displacive degrees of freedom than those strictly needed to describe the tilting of the octahedra. Using these models for fitting scattering data leads to structures where the octahedra are distorted in a way that cannot be represented in terms of the pure Glazer tilt patterns with rigid units even in the cases where the octahedra are not geometrically required to distort \cite{howardGroupTheoreticalAnalysisOctahedral1998}.

Here we explore a more direct approach to modeling collective rotations by using algebraic expressions that link displacements of atoms in the Glazer tilt systems.  Going beyond purely symmetry constraints is surprisingly challenging.  Approximate Monte Carlo approaches have been attempted \cite{sartbaevaprlgeometric2006,sartbaevaQuadrupolarOrderingLaMnO2007d}, where atoms are tethered to rigid-unit templates which do not distort, but are allowed to relax away from the vertices.  It has also been shown \cite{campbellAlgebraicApproachCooperative2018} that for small rotations a set of linear equations on top of symmetry mode analysis \cite{perez-matoModeCrystallographyDistorted2010a} can identify collective modes in a system of connected rigid units that do not (or hardly) distort the units. However, there is currently no straightforward way of incorporating this information into a refinement program for quantitative modeling of data in terms of this collective mode basis.
    
Our approach of explicitly building the geometric constraint equations without assumed symmetries has an advantage that it can be easily plugged into local structure modeling schemes such as that used in the diffpy-CMI \cite{juhasComplexModelingStrategy2015e} program. The program works in the P1 space-group by design, allowing one to introduce structural distortions by moving atoms at will. The approach greatly reduces the number of refinable parameters in a physically meaningful way and can help to build intuition about the structure and how it is likely to distort. It also allows the user to test directly hypotheses about the rigidity of the units or the type of tilting present in a sample without the conceptual complexity of having to surf between space groups. This can give new insight that might be lost otherwise. Here we demonstrate the use of our code on the compound \ch{CaTiO3}, the archetypal perovskite with a well-known Glazer tilt pattern $\alpha^+ \beta^- \beta^-$.

\section{Glazer tilt definitions}

The Glazer tilt systems, as laid out by Glazer in 1972 \cite{glazerClassificationTiltedOctahedra1972a}, describe the complete set of collective rotations allowed in a network of corner-shared octahedra as found in perovskites\footnote{The tilt patterns described by Glazer can all be described using a 2x2x2 (or smaller) supercell of the cubic perovskite unit cell, and collective distortions requiring larger supercells are unlikely.}, shown in Table~\ref{tab:glazersystems}. 

\begin{table*}[bh]
    \centering
    \caption{For each of the different Glazer tilt patterns we give the index as assigned by Glazer (Glazer, A.~M.~1972; Glazer, A.~M.~1975), the tilts given with Glazer notation and the space group symmetry of the resulting phase. Note that we have only included here the tilt systems that are symmetry nonequivalent.}
    \begin{tabular}{|p{1.5cm} p{1.7cm} p{3cm}|}
    \hline
    Tilt \\ system & Tilts & Space group\\ [0.3ex] 
    \hline\hline
    23 & $\alpha^0 \alpha^0 \alpha^0$ & $Pm\Bar{3}m$ (\#221) \\
    \hline
    22 & $\alpha^0 \alpha^0 \gamma^-$ & $I4/mcm$ (\#140) \\
    \hline
    21 & $\alpha^0 \alpha^0 \gamma^+$ & $P4/mbm$ (\#127) \\
    \hline
    20 & $\alpha^0 \beta^- \beta^-$ & $Imma$ (\#74) \\
    \hline
    19 & $\alpha^0 \beta^- \gamma^-$ & $C2/m$ (\#12) \\
    \hline
    17 & $\alpha^0 \beta^+ \gamma^-$ & $Cmcm$ (\#63)  \\
    \hline
    16 & $\alpha^0 \beta^+ \beta^+$ & $I4/mmm$ (\#139) \\
    \hline
    14 & $\alpha^- \alpha^- \alpha^-$ & $R\Bar{3}c$ (\#167) \\
    \hline
    13 & $\alpha^- \beta^- \beta^-$ & $C2/c$ (\#15) \\
    \hline
    12 & $\alpha^- \beta^- \gamma^-$ & $P\Bar{1}$ (\#2) \\
    \hline
    10 & $\alpha^+ \beta^- \beta^-$ & $Pnma$ (\#62) \\
    \hline
    8 & $\alpha^+ \beta^- \gamma^-$ & $P2_1/m$ (\#11) \\
    \hline
    5 & $\alpha^+ \alpha^+ \gamma^-$ & $P4_2/nmc$ (\#137) \\
    \hline
    3 & $\alpha^+ \alpha^+ \alpha^+$ & $Im\Bar{3}$ (\#204) \\
    \hline
    1 & $\alpha^+ \beta^+ \gamma^+$ & $Immm$ (\#71)  \\ [1ex] 
    \hline
    \end{tabular}
    
    \label{tab:glazersystems}
\end{table*}
For the sake of clarity, we introduce the naming scheme here. An octahedron can be tilted around one, two or all three of the cartesian axes, $x$, $y$ and $z$. The nature of each rotation is indicated by three Greek letters with superscripts, where the first letter denotes the rotation around $x$, the second around $y$ and the third around $z$. Repeating letters (e.g. $\alpha^+ \alpha^+ \alpha^+$) indicate that the amplitudes around the specific axes are the same, while different letters (e.g. $\alpha^+ \beta^+ \gamma^+$) indicate that the tilts differ in amplitude around the different axes. 

The superscripts can take the value 0, + or - to indicate a zero-tilt amplitude or a non-zero amplitude with tilts in adjacent layers along the tilt axis being either in-phase (+) or out-of-phase (-). For example, the tilt pattern $\alpha^0 \alpha^0 \gamma^+$ has no tilt around the $x$ and $y$ axes and a non-zero tilt around the $z$ axis. Because of the connectivity of the octahedra at their corners, neighbouring octahedra in the plane perpendicular to the tilt axis rotate in the opposite direction to the central octahedron, leading to a doubling of the unit cell in that plane. In the example of $\alpha^0 \alpha^0 \gamma^+$, the unit cell is therefore doubled in the $ab$ plane, but not along the $z$ axis. On the other hand, an out-of-phase tilt, for example along the $z$ axis in the pattern $\alpha^0 \alpha^0 \gamma^-$, will double the unit cell also along the tilt axis. \fig{illustration-of-IP-OOP-tilts} illustrates the difference between the in-phase and out-of-phase tilt pattern of the $\alpha^0 \alpha^0 \gamma^+$ and $\alpha^0 \alpha^0 \gamma^-$ tilt systems, as viewed down the tilt axis.

\section{Approach}

Here we describe the method for building constrained Glazer tilt pattern models. The code may be found at https://github.com/sandraskj/glazer\_fitting.

Models are built using the diffpy-CMI program \cite{juhasComplexModelingStrategy2015e}, which has powerful and flexible methods for specifying constraints between model parameters. This allows, in principle, large numbers of parameters to be expressed in terms of a much smaller number of variables from analytic or numerical expressions. We first generate the constraints as symbolic expressions relating multiple atoms' fractional coordinates. These expressions are then captured into the diffpy-CMI constraint handling interface.

For all the Glazer tilt systems listed in Table~\ref{tab:glazersystems} the constraints have been constructed such that the shortest B-X distances are all kept rigid. Since, for most of the systems, there is a small coupling between the rotation modes around the three axes, these constraints will lead to a small octahedral distortion and octahedral angles deviating slightly from 90$^{\circ}$, so the tilts are not strictly rigid.  However, the tilt equations result in almost rigid octahedral tilting, where the collective modes may be described just in terms of tilt angles around each axis which are the only refinable parameters for the modes when fitting to data, in addition to the cubic lattice parameter. 

The collective octahedral rotations do not include A-site ion structural parameters. Although the A atoms are not directly part of the octahedral tilting network,  their positions are still refined, as they do respond to the tilts by displacing. We chose to constrain the A cation displacements in such a way as to respect the expected symmetry of the tilted structure, which in the case of \ch{CaTiO3} is the space group $Pnma$. 

Activating tilt modes leads to well-defined reductions in the lattice parameters, and therefore for a full description, we need to find the appropriate scaling parameters expressed in terms of the Glazer tilt amplitudes and the lattice parameter of the cubic parent structure. We start with the interatomic vectors from the B atom at the origin to its three unique X neighbors in the octahedron, $\vec{r}_{\mathrm{B-X1}}$, $\vec{r}_{\mathrm{B-X2}}$ and $\vec{r}_{\mathrm{B-X3}}$,
\begin{align}
    |\vec{r}_\mathrm{B-X1}| &= \sqrt{ \left(a'\frac{1}{4} \right)^2 + \left(b' \cdot y_\mathrm{X1}\right)^2 + \left( c' \cdot z_\mathrm{X1}\right)^2}\, ,\label{eq:O1vec}\\
    |\vec{r}_\mathrm{B-X2}| &= \sqrt{\left( a'\cdot x_\mathrm{X2}\right)^2 + \left( b' \frac{1}{4} \right)^2 + \left( c' \cdot z_\mathrm{X2} \right)^2}\, ,\\
    |\vec{r}_\mathrm{B-X3}| &= \sqrt{\left( a'\cdot x_\mathrm{X3}\right)^2 + \left( b' \cdot y_\mathrm{X3}\right)^2 + \left( c' \frac{1}{4} \right)^2}\, , \label{eq:O3vec}
\end{align}
where $a'$, $b'$ and $c'$ are the lattice parameters of the distorted supercell for a given set of Glazer tilts. Keep in mind that the fractional coordinates $y_{\mathrm{X1}}$, $z_{\mathrm{X2}}$ etc., are all expressions containing the Glazer tilt variables and the lattice parameter of the cubic parent structure $a_h$.
Next, we set each of the bond lengths to be a quarter of the parent unit cell $a_h$, i.e., for X1,
\begin{align}
    |\vec{r}_{\mathrm{B-X1}_x}| = |\vec{r}_{\mathrm{B-X1}_y}| = |\vec{r}_{\mathrm{B-X1}_z}| = a_h/4\,,
\end{align}
where $\vec{r}_{\textrm{B-X1}_z}$ is the $z$ component of $\vec{r}_\textrm{B-X1}$.  Since the rotations are assumed to be rotations of rigid octahedra, these lengths will not change after the rotation.
This allows us to relate the lattice parameter of the Glazer tilt distorted supercell to the ones of the cubic parent cell through scaling parameters $s_a$, $s_b$ and $s_c$,
\begin{align}
    a' = \frac{2a_h}{s_a}\,,\;  b' = \frac{2a_h}{s_b}\,,\;  c' = \frac{2a_h}{s_c}.
\end{align}
Substituting for $a'$, $b'$ and $c'$ in Eqs.~\ref{eq:O1vec}-\ref{eq:O3vec}, we get a set of three equations,
\begin{align}
    |\vec{r}_\mathrm{B-X1}| = \sqrt{\frac{1}{16s_a} + \left(\frac{y_\mathrm{X1}}{s_b}\right)^2 + \left(\frac{z_\mathrm{X1}}{s_c}\right)^2} = a_h/4 \, ,\\ 
    |\vec{r}_\mathrm{B-X2}| = \sqrt{\left( \frac{x_\mathrm{X2}}{s_a} \right)^2 + \frac{1}{16s_b} +\left( \frac{z_\mathrm{X2}}{s_c} \right)^2} = a_h/4 \,,\\
    |\vec{r}_\mathrm{B-X3}| = \sqrt{ \left( \frac{x_\mathrm{X3}}{s_a} \right)^2 + \left(\frac{y_\mathrm{X3}}{s_b} \right)^2 + \frac{1}{16s_c}} = a_h/4\,,
\end{align}
 that can be solved for $s_a$, $s_b$ and $s_c$,
\begin{align}
    s_a &= \sqrt{\frac{  1- 256 x_\mathrm{X2}^2 \left( y_\mathrm{X1}^2 -16 y_\mathrm{X3}^2 z_\mathrm{X1}^2 \right) - 256 \left( y_\mathrm{X3}^2 z_\mathrm{X2}^2 + x_\mathrm{X3}^2 \left( z_\mathrm{X1}^2 - 16 y_\mathrm{X1}^2 z_\mathrm{X2}^2\right)\right)  }
    {  1 - 16 y_\mathrm{X1}^2 + 16 \left(-1+16 y_\mathrm{X3}^2\right) z_\mathrm{X1}^2 + 256 (y_\mathrm{X1}^2-y_\mathrm{X3}^2) z_\mathrm{X2}^2  }}\,, \\
    s_b &= \sqrt{\frac{  1- 256 x_\mathrm{X2}^2 \left( y_\mathrm{X1}^2 -16 y_\mathrm{X3}^2 z_\mathrm{X1}^2 \right) - 256 \left( y_\mathrm{X3}^2 z_\mathrm{X2}^2 + x_\mathrm{X3}^2 \left( z_\mathrm{X1}^2 - 16 y_\mathrm{X1}^2 z_\mathrm{X2}^2\right)\right)  }
    {  1 + 16 x_\mathrm{X2}^2 \left(-1 + 16 z_\mathrm{X1}^2\right) - 16 z_\mathrm{X2}^2 + 256 x_\mathrm{X3}^2 \left(-z_\mathrm{X1}^2 + z_\mathrm{X2}^2\right)  }}\,, \\
    s_c &= \sqrt{\frac{  -1- 256 x_\mathrm{X2}^2 \left( y_\mathrm{X1}^2 -16 y_\mathrm{X3}^2 z_\mathrm{X1}^2 \right) + 256 \left( y_\mathrm{X3}^2 z_\mathrm{X2}^2 + x_\mathrm{X3}^2 \left( z_\mathrm{X1}^2 + 16 y_\mathrm{X1}^2 z_\mathrm{X2}^2\right)\right)  }
    {  -1 + x_\mathrm{X3}^2 (16-256 y_\mathrm{X1}^2) + 256 x_\mathrm{X2}^2  (y_\mathrm{X1}^2 - y_\mathrm{X3}^2) + 16 y_\mathrm{X3}^2  }}\,.
\end{align}
Setting these as constraints in the refinement allows the unit cell to change size according to the tilt amplitude without introducing any extra refinable parameters.

We present the full constraints for Glazer system 10, as generated from the code, in the project code repository on GitHub (https://github.com/sandraskj/glazer\_fitting).
In the GitHub repository we also provide the code that generates the constraints for all the Glazer systems and brief instructions for how the reader can download them and how to set it up for their own refinements using diffpy-CMI.

\section{Experimental measurements}

To obtain experimental pair distribution functions for \ch{CaTiO3} measurements were carried out at the 28-ID-2 (XPD) beamline at the NSLS-II at Brookhaven National Laboratory on a commercially purchased powder sample of \ch{CaTiO3} (Strem Chemicals Inc, CAS 12049-50-2). 
A 2D Perkin Elmer amorphous silicon detector was placed 380~mm behind the sample, which was loaded in a 0.5~mm glass capillary. The wavelength of the incident x-rays was 0.16635~\AA. Data were collected at 200~K for 60~s in a flowing nitrogen cryostream.

The data were processed using standard methods \cite{egamibookunderneaththebraggpeak2012}. The instrument geometry was calibrated using data from a fine powdered Ni powder using pyFAI \cite{kiefferNewToolsCalibrating2020b}.  
2D diffraction patterns were processed by applying masks to remove the beam stop as well as outlier saturated and dead pixels using a home-written automasking protocol.  After correction for polarization effects they were integrated azimuthally along circles of constant scattering vector magnitude, $Q$, also using pyFAI. The background signal from an empty glass capillary was subtracted and the data were normalized and corrected to obtain the reduced total scattering structure function, $F(Q)$, which was Fourier transformed to obtain the PDF.  This was done using  PDFgetX3~\cite{juhasPDFgetX3RapidHighly2013h}. The maximum range of data used in the Fourier transform was $Q_{max} = 23.6$~\AA$^{-1}$\

\section{Results}

Our initial tests of the approach are carried out on simulated data from the known ground-state structure of \ch{CaTiO3}.
\begin{figure*}
    \centering
    \includegraphics[width=0.7\textwidth]{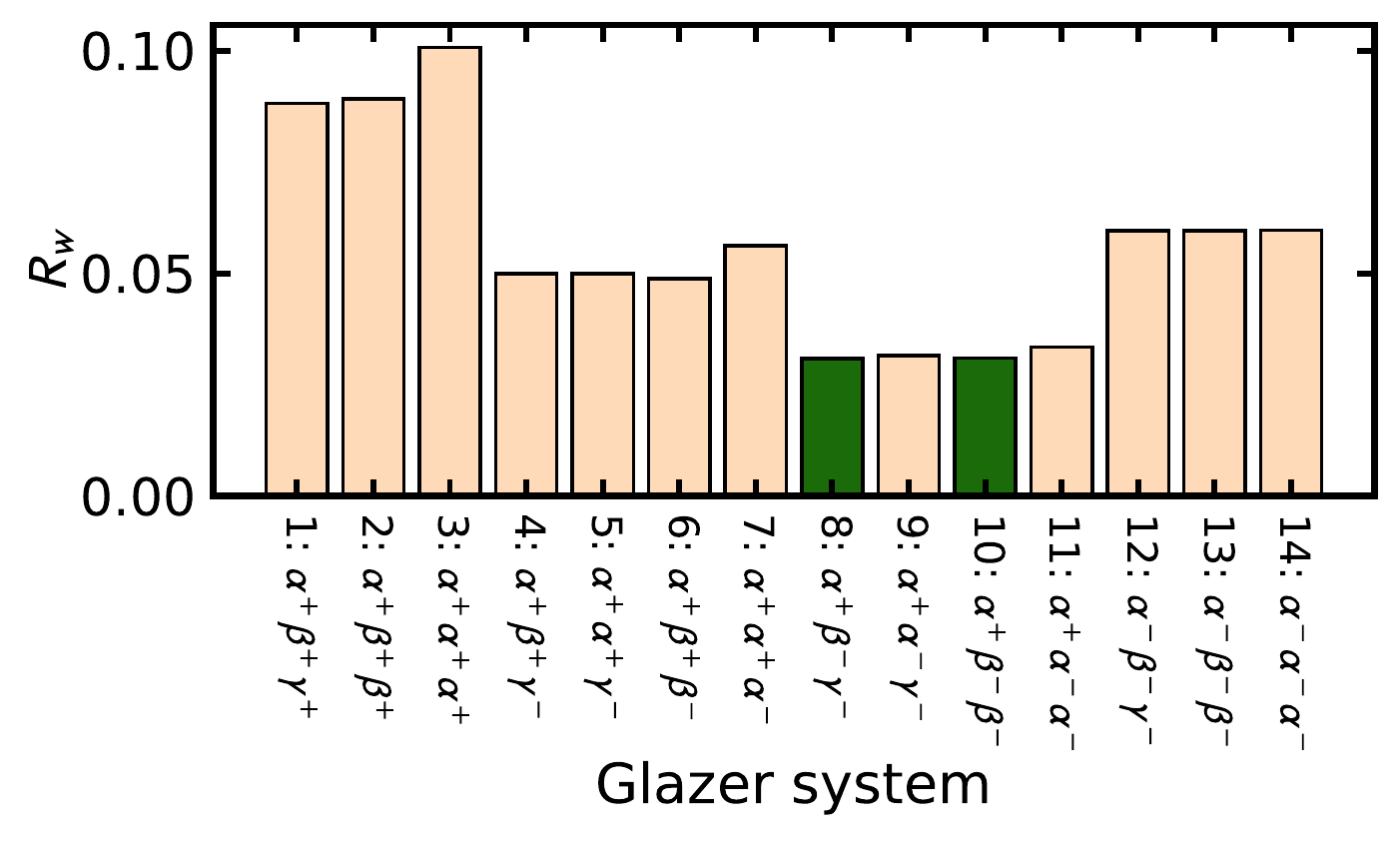}
    \caption{Comparison of the fit of the 14 three-tilt Glazer systems to a calculated PDF of \ch{CaTiO3} with octahedral rotations but without Ca displacements.}
    \label{fig:comparing-14-tilt-systems}
\end{figure*}
The structure was created in Glazer system no.~10 with an in-phase tilt around one axis of $\alpha=9^{\circ}$ and out-of-phase tilt around the other two axes of $\beta=10^{\circ}$. For simplicity the simulated structure had no displacement of the Ca atoms away from their cubic positions. The isotropic atomic displacement parameters (ADPs) for all the ions were set to $U_{iso}(\mathrm{Ca}) = 0.0030$~\AA$^{2}$, $U_{iso}(\mathrm{Ti}) = 0.0046$~\AA$^{2}$, and $U_{iso}(\mathrm{O}) = 0.011$~\AA$^{2}$, similar to those obtained from fitting \ch{CaTiO3} with a conventional $Pnma$ model.

The PDF was calculated from the structure using diffpy-CMI \cite{juhasComplexModelingStrategy2015e}, with damping and broadening parameters set to values 0.029~\AA$^{-1}$ and 0.010~\AA$^{-1}$, respectively, obtained from the calibration sample in our measurement and $Q_{max} = 23.6$~\AA$^{-1}$, the same value as we used for the experimental data.
 
We then fit constrained Glazer models from each of the 22 Glazer tilt patterns to the data from $1.6<r<15$~\AA\ to see how well the constrained refinements were working. The starting values for the tilt amplitudes in the refinements models were set to values that were roughly 70\% of the true values in the structure for the calculated data set.
  
The fits with the one-tilt and two-tilt models were poor in most cases, while all the three-tilt systems gave fit residuals below 10\%. Tilt system 10 (the ground-truth result) is one of the three-tilt systems so this gives confidence that the approach can easily differentiate the presence or absence of tilts.  However, within the subset of three-tilt systems, different families of tilt combinations can be found which refine to significantly different \rw  values, as shown in \fig{comparing-14-tilt-systems}.  Interestingly, the fits can differentiate cases that have $+++$, $++-$, $+--$ and $---$ tilts, but within those families it cannot distinguish between different Glazer systems. This may be because the tilt amplitudes we chose for the test, coming from the observed values in \ch{CaTiO3}, are close to each other. 

The best overall fit was found for Glazer system 10, the correct one, as well as Glazer system 8 that has the same tilt pattern, but with an extra degree of freedom that allows the out-of-phase tilts to be of different amplitudes. This shows that the collective mode refinements are working in diffpy-CMI.

\begin{table*}[]
    \centering
        \caption{Comparison of parameters from the space group and Glazer model refinements over the $r$ range 1.6-50 \AA. The parameters typeset in bold were variables that were explicitly refined, while the parameters typeset in italic were extracted or calculated post-refinement from the refined values. Two values are given each for $\alpha$ and $\beta$ in the space group model because different octahedra tilt by different amounts.}
    \begin{tabular}{c  c | c  c}
         \multicolumn{2}{c|}{$Pnma$ space group model} & \multicolumn{2}{c}{Glazer model} \\
         variable & value & variable & value   \\
         \hline \hline
         \textbf{scale} & 0.18  & \textbf{scale} & 0.17   \\
         \textbf{delta1} & 1.03  & \textbf{delta1} & 2.47   \\
         \hline
             &        & $\mathbf{a_h}$  & 3.907  \\
         $\mathbf{a}$ & 5.428  & $a$ &  5.402   \\ 
         $\mathbf{b}$ & 7.620  & $b$ &  7.594   \\
         $\mathbf{c}$ & 5.366  & $c$ &  5.402   \\
         \hline
         $\mathbf{x_{Ca}}$ & 0.0357 & $\mathbf{x_{Ca}}$ & 0.0216\\
         $\mathbf{z_{Ca}}$ & 0.0031 & $\mathbf{y_{Ca}}$ & 0.0069\\
         \hline
         $\alpha$ & 9.6$^{\circ}$, 10.1$^{\circ}$  & $\boldsymbol{\alpha}$ & 7.6$^{\circ}$  \\
         $\beta$  & 7.1$^{\circ}$, 10.6$^{\circ}$  & $\boldsymbol{\beta}$  & 9.7$^{\circ}$  \\
         $\mathbf{x_{O1}}$ & 0.2059  &  &    \\
         $\mathbf{y_{O1}}$ & 0.0335  &  &    \\
         $\mathbf{z_{O1}}$ & 0.2073  \\
         $\mathbf{x_{O2}}$ & 0.0155  \\
         $\mathbf{z_{O2}}$ & 0.5784  \\
         \hline
         $\mathbf{U_{iso}}$\textbf{(Ca)} & 0.005 & $\mathbf{U_{iso}}$\textbf{(Ca)} & 0.004   \\
         $\mathbf{U_{iso}}$\textbf{(Ti)} & 0.003 & $\mathbf{U_{iso}}$\textbf{(Ti)} & 0.004 \\
         $\mathbf{U_{iso}}$\textbf{(O)} & 0.010 & $\mathbf{U_{iso}}$\textbf{(O)} & 0.011 \\
         \hline
         \rw & 0.087 & \rw & 0.245    \\
    \end{tabular}
    \label{tab:variables}
\end{table*}

Next, we performed refinements on an experimental data set of \ch{CaTiO3}. We performed the refinements with two models: One using our formulation based on Glazer tilt system 10 and for comparison, a model with constraints consistent with the crystallographic space-group $Pnma$, which does not impose rigid tilts. The Ca sites were constrained the same way in both models, according to the space group symmetry of $Pnma$.  The space group model has~10 structural degrees of freedom, while the Glazer model has only~5. The variables, including explicitly refined as well as post-calculated ones, and their values after refinement over 1.6-50~\AA\ are listed in Table~\ref{tab:variables}.

\begin{figure*}
    \centering
    \includegraphics[width=0.7\textwidth]{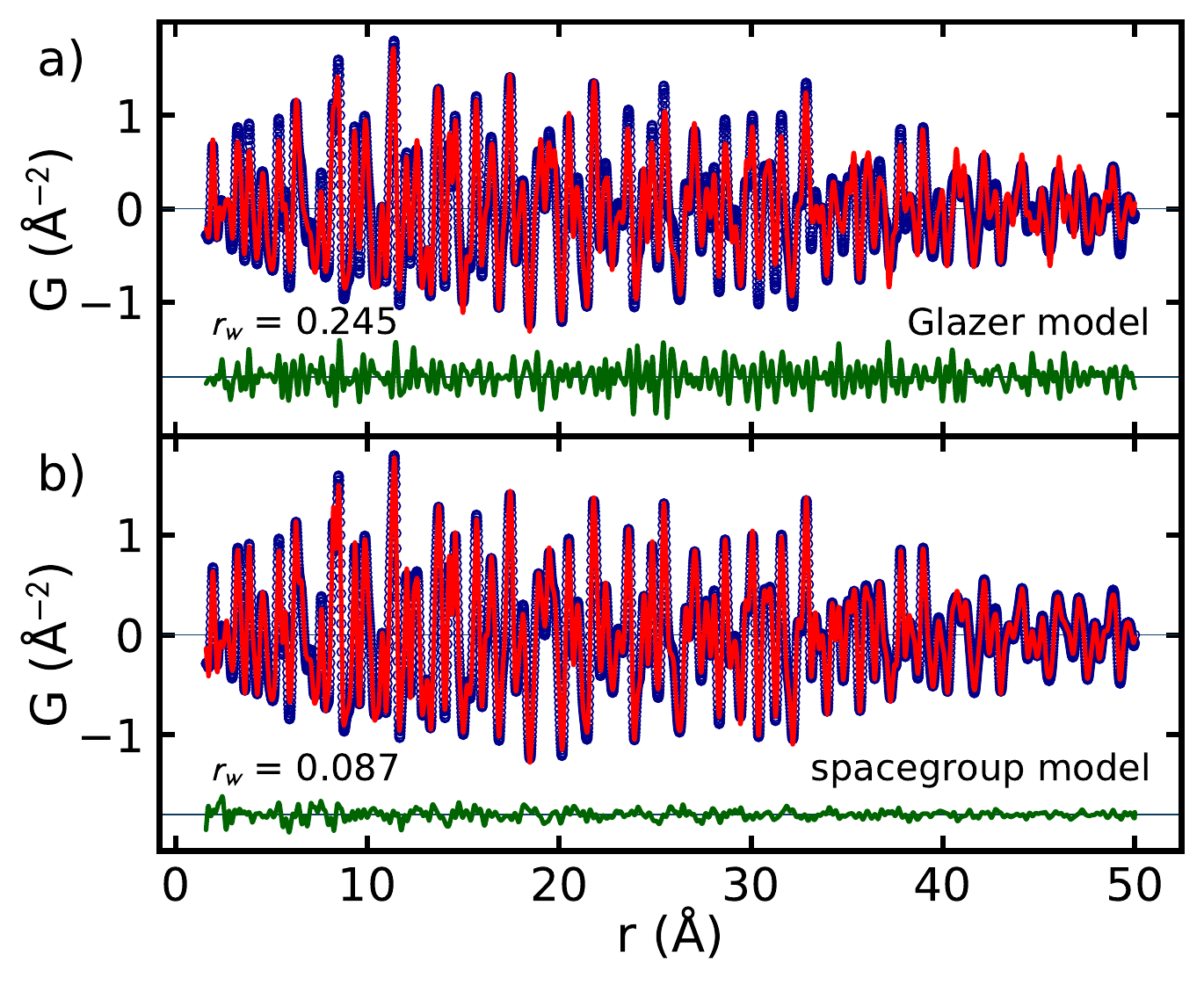}
    \caption{Plots of measured (blue) and best-fit (red) PDFs of \ch{CaTiO3} with the difference curve plotted in green offset below over the $r$ range 1.6-50Å. The model for the best-fit PDF is from (a) the constrained Glazer tilt model in Glazer system 10 and (b) allowing all the structural degrees of freedom from the $Pnma$ structural model. 
    }
    \label{fig:fitcurves-glazer-spacegroup_1-50AA}
\end{figure*}

Fitting both models over this wide $r$ range (\fig{fitcurves-glazer-spacegroup_1-50AA}), we can see that the space group model gives a significantly smaller fit residual (space group gives \rw = 0.087 while the Glazer model gives \rw = 0.245), which is not surprising given its larger number of refinable variables. Comparing the refined structural parameters from the two models, we see that all, except for the tilt angles and the lattice parameters are in quite good agreement as shown by comparing the values in Table~\ref{tab:variables}.
The information of interest to us is the presence and amplitude of rigid Glazer tilt modes.  For the Glazer model these are a direct output of the program.  For comparison, we also calculate the tilt angles from the space group by finding the angle that each vector between opposite pairs of oxygen atoms on an octahedron makes with the pseuodocubic axes. Previous studies have calculated the angles using the oxygen positions similarly to what we do here, but in such a way that gives the average over two octahedra in order to to end up with multiple values for each Glazer component \cite{kennedyPhaseTransitionsPerovskite1999, yashimaStructuralPhaseTransition2009}. This is necessary if you want a single value for each tilt component as the space group model does not keep the octahedral bond lengths rigid, and the tilts therefore vary depending on the bond chosen for evaluating it. For the sake of comparing the performance of our Glazer model to the space group model, we choose here to not present such average tilt angles, but rather present the tilts are they present themselves on different octahedra. This will highlight the difference in rigidity and robustness of the two models.

The Glazer model results in values of in-phase tilt $\alpha = 7.6^{\circ}$ and out-of-phase tilt $\beta = 9.7^{\circ}$. The space group model gives the values $\alpha = 9.6^{\circ}$ and 10.1$^{\circ}$ and $\beta = 7.1^{\circ}$  and 10.6$^{\circ}$.  The average value of the in-phase $\alpha$ tilt is higher in the space-group model by almost 2$^{\circ}$ than in the Glazer model, and that of the out-of-phase $\beta$ tilt is lower by about 1$^{\circ}$.  In the case of the $\beta$ the two different octahedra in the space group model are quite different and actually straddle the value obtained in the Glazer fit.

Addressing the difference in fit quality, we believe that a significant contribution to the poorer fit is due to the tighter constraints on the lattice parameters in the Glazer fits. The space group model is orthorhombic, with three different lattice parameters.  For the Glazer model only one lattice parameter variable is refined.  The tilts then result in a change in shape of the unit from cubic, as described above.  We note that the particular tilt pattern in this Glazer mode results in a tetragonal, not orthorhombic, unit cell (Table~\ref{tab:variables}). The $a = c$ parameters for the Glazer model lie in value between those of the space-group model, but are not able to separate into short and long values allowed by the orthorhombic crystallographic model due to the Glazer model constraints, whereas clearly structural relaxations beyond the rigid tilts are present in the actual material that prefer this.   

\begin{figure*}
    \centering
    \includegraphics[width=0.7\textwidth]{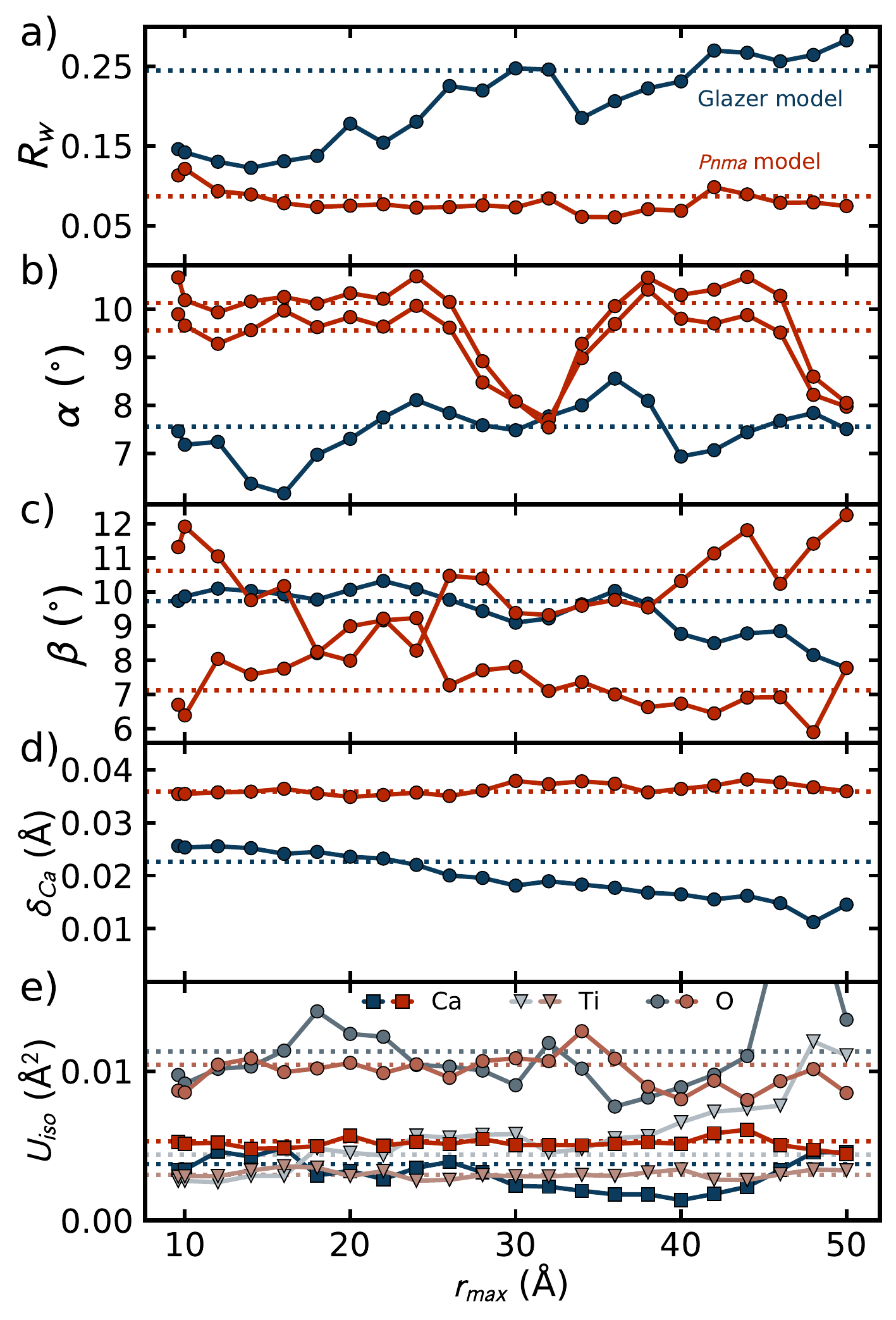}
    \caption{Comparison of the a) fit residual~\rw, the octahedral tilt amplitudes b) $\alpha$ and c) $\beta$, d) the total Ca displacements ($\delta_{Ca}$) e) and the $U_\mathrm{iso}$ values from boxcar fits with the space group model and the Glazer model of \ch{CaTiO3} at 200~K. The $r$ range (or 'the box') was set to 8~\AA\ and incrementally shifted to higher $r$ values in steps of 2~\AA. The labels on the $x$-axis correspond to the highest value in the box, $r_{max}$. The dotted lines represent the values obtained from a fit over the 1.6-50 \AA\ range. We note that for the space group model, the tilt angles $\alpha$ and $\beta$ differ depending on which octahedra were used to calculate them, and such are represented by two different lines.
    }
    \label{fig:boxcar_comparison}
\end{figure*}

If the difference in \rw between the two models in the wide-range fits is due to the difference in model rigidity, the models would be expected to perform more comparably when fitting only the most local structure, and for the Glazer model to perform worse at higher values of $r$. The Glazer model only allows for the degrees of freedom that are strictly necessary for the tilt pattern $\alpha^+ \beta^- \beta^-$, and comparing the two fits at different length scales therefore allows us to separate contributions to the PDF signal that come from rigid tilts and additional non-rigid relaxations. It is also interesting to consider if the refined values of the Glazer tilts varies with the $r$-range that is fit over, as might be the case if the tilts become damped with increasing-$r$. We therefore performed a series of fits where an $r$-range of a fixed size (referred to as a box) is shifted incrementally up to higher values, an approach we call a `boxcar' fit. The $r$-dependence of the refined variables are shown in \fig{boxcar_comparison}.

As evident in \fig{boxcar_comparison}(b) and (c) the values of the tilt amplitudes vary more smoothly in the Glazer model than in the space-group model indicating that refinement of these variables is more stable in the more highly constrained Glazer fits. Also, whilst the $\alpha$ tilt is fairly $r$-independent, there is a marked tendency for the $\beta$ tilt to decrease with increasing-$r$ in the Glazer fit. This would be expected if there is a loss in the coherence of the tilts with increasing-$r$ due to a non-rigidity.  This suggests that in cases where local tilts survive in a material but are not present globally, the range of coherence of the collective motions may be measured by this approach.  We also see a similar trend in the total displacement of Ca from cubic positions ($\delta_{Ca}$, panel d), with the Glazer model trending downwards, while the space group model stays at the same value throughout the $r$ range.

\begin{figure*}
    \centering
    \includegraphics[width=0.7\textwidth]{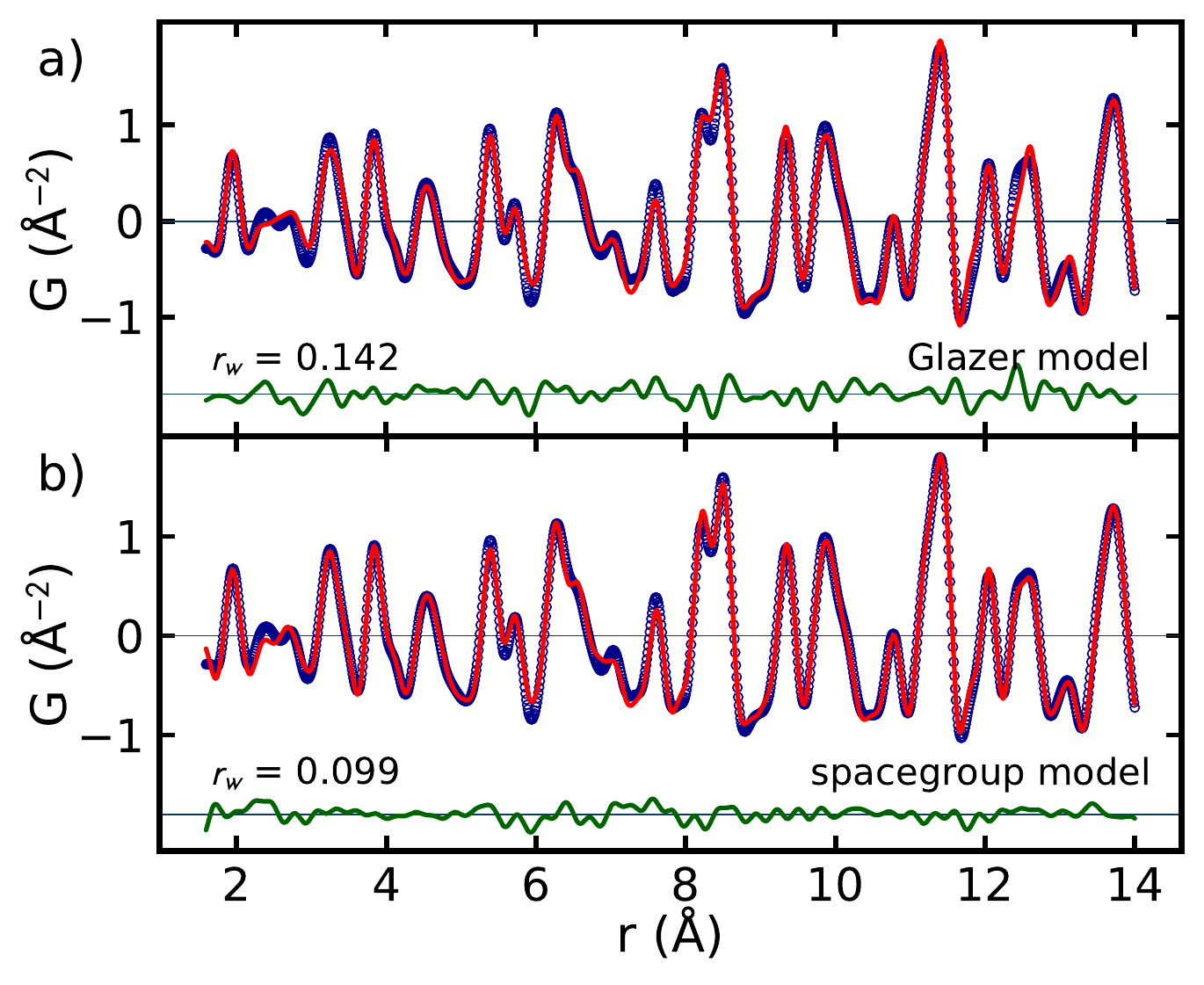}
    \caption{Plots of measured (blue) and best-fit (red) PDFs of \ch{CaTiO3} with the difference curve plotted in green offset below over an $r$ range of 1.6-14 Å. The model for the best-fit PDF is from (a) the constrained Glazer tilt model in Glazer system 10 and (b) allowing all the structural degrees of freedom from the $Pnma$ structural model. }
    \label{fig:fitcurves-glazer-spacegroup_1-14AA}
\end{figure*}

Since we believe that a non-rigidity in the tilts explains the difference in behavior of the two fits, a comparison for the fits over a much narrower range $r=1.6-14$~\AA\ we might expect much better agreement between the Glazer and space-group models.  This is indeed found in the comparable \rw values at low-$r$ (\fig{boxcar_comparison}(a)).  We show the fits over this low-$r$ region on an expanded scale in  \fig{fitcurves-glazer-spacegroup_1-14AA}. This shows that the most local structure is sufficiently rigid, with octahedral tilts being the predominant structural distortion, such that it can be well represented by the Glazer model.

We note that for the case we studied, \ch{CaTiO3} at room temperature, the tilts are long-range ordered and so are expected to persist over large distances, asymptotically approaching the crystallographic values. This kind of boxcar analysis can be expected to be more interesting in materials where no tilts are observed in the average structure but are observed locally~\cite{skjaervoUnconventionalContinuousStructural2019a,bozin;nc19,kochDualOrbitalDegeneracy2021,yangtwoorbitaldegeneracyprb2020, wangUnderstandingElectronicPeculiarities2020a,sennEmergenceLongRangeOrder2016i}.

\section{Conclusions}

We have developed sets of constraint equations that explicitly model octahedral tilts (Glazer tilts) in perovskites. The model allows refinements of collective atomic motions by geometrically connecting atoms in the lattice allowing rigid rotations to be modeled directly. We have implemented the constraints directly in the PDF modeling program diffpy-CMI.

We  have demostrated the use of our code on the canonical tilted perovskite system \ch{CaTiO3}, which has a known long-range ordered Glazer tilt system $\alpha^+ \beta^- \beta^-$. We found that our Glazer model fits comparably to the known space group model $Pnma$ below $r = 14$~\AA. We further saw that the Glazer model performed progressively worse at higher $r$, due to the rigidity of the model.  In this case the rigid tilts alone broke the cubic symmetry to tetragonal, whereas the observed symmetry is orthorhombic, which explains the discrepancy in the fit residuals.  Presumably, non-rigid relaxations and relaxations of atoms not involved in the tilts are responsible for the additional reduction in symmetry.

The use of our Glazer model could be used to study a wide range of perovskite systems to better understand whether their structure is well explained in terms of pure octahedral rotations, how the rotations vary with parameters such as temperature and pressure, and what additional structural relaxations are needed to explain the structure beyond the simple picture of octahedral rotations.  The highly constrained fits can be expected to give stable refinements even when data quality is limited, for example, from small nanoparticles or powders in a diamond anvil cell.  The work also highlights the strengths and limitations of the geometric approach in building rigid body constraints.

\section{Acknowledgements}
We acknowledge Daniel E.~DeRosha and Jonathan Owen for help in preparing the \ch{CaTiO3} sample, and Milinda Abeykoon and Gihan Kwon for help with collecting data at the 28-ID-1 (PDF) beamline at the NSLS-II.  
This work was supported by the U.S. National Science Foundation through grant DMREF-1922234. M.A.K. acknowledges support from the Carlsberg Foundation (grant no. CF17-0823).  Use of the National Synchrotron Light Source II, Brookhaven National Laboratory, was supported by the US Department of Energy, Office of Science, Office of Basic Energy Sciences, under Contract No. DE-SC0012704.


\end{document}